\documentclass[prd, notitlepage, floatfix, print, onecolumn, showpacs, showkeys, nofootinbib, superscriptaddress]{revtex4-1}
\usepackage{natbib}
\usepackage{times}
\usepackage{amssymb,amsbsy,amsmath,amsfonts}
\usepackage{graphicx}
\usepackage{float}
\usepackage{color}
\usepackage{morefloats}
\usepackage{rotating}
\usepackage{srcltx}
\usepackage{slashed}
\usepackage{subfigure}
\usepackage{multirow}
\usepackage{verbatim}
\usepackage{hyperref}
\usepackage{tabularx}

\usepackage{epstopdf}

\graphicspath{{./}{./img/}{./fig/}{./image/}{./figure/}{./picture/}}

\begin{document}

\title{Consistency between SU(3) and SU(2) covariant baryon chiral perturbation theory for the nucleon mass}

\author{Xiu-Lei Ren}
\affiliation{School of Physics and
Nuclear Energy Engineering \& International Research Center for Nuclei and Particles in the Cosmos \&
Beijing Key Laboratory of Advanced Nuclear Materials and Physics,  Beihang University, Beijing 100191, China}
\affiliation{State Key Laboratory of Nuclear Physics and Technology, School of Physics, Peking University, Beijing 100871, China}

\author{L. Alvarez-Ruso}
\affiliation{Instituto de F\'isica Corpuscular (IFIC),
Centro Mixto Universidad de Valencia-CSIC, Institutos de Investigaci\'{o}n de Paterna,
Apartado 22085, 46071 Valencia, Spain}

\author{Li-Sheng Geng}
\affiliation{School of Physics and
Nuclear Energy Engineering \& International Research Center for Nuclei and Particles in the Cosmos \&
Beijing Key Laboratory of Advanced Nuclear Materials and Physics,  Beihang University, Beijing 100191, China}

\author{Tim Ledwig}
\affiliation{Departamento de F\'{\i}sica Te\'{o}rica and IFIC,
Centro Mixto Universidad de Valencia-CSIC, Institutos de Investigaci\'{o}n de Paterna,
Apartado 22085, 46071 Valencia, Spain}

\author{Jie Meng}
\affiliation{School of Physics and
Nuclear Energy Engineering \& International Research Center for Nuclei and Particles in the Cosmos, Beihang University, Beijing 100191, China}
\affiliation{State Key Laboratory of Nuclear Physics and Technology, School of Physics, Peking University, Beijing 100871, China}
\affiliation{Department of Physics, University of Stellenbosch, Stellenbosch 7602, South Africa}

\author{M. J. Vicente Vacas}
\affiliation{Departamento de F\'{\i}sica Te\'{o}rica and IFIC,
Centro Mixto Universidad de Valencia-CSIC, Institutos de Investigaci\'{o}n de Paterna,
Apartado 22085, 46071 Valencia, Spain}

\begin{abstract}
Treating the strange quark mass as a heavy scale compared to the light quark mass,
we perform a matching of the nucleon mass in the SU(3) sector to the two-flavor case in covariant baryon chiral perturbation theory. The validity of the $19$ low-energy constants appearing in the octet baryon masses up to next-to-next-to-next-to-leading order~\cite{Ren:2014vea} is supported by comparing the effective parameters (the combinations of the $19$ couplings) with the corresponding low-energy constants in the SU(2) sector~\cite{Alvarez-Ruso:2013fza}. In addition, it is shown that the dependence of the effective parameters and the pion-nucleon sigma term on the strange quark mass is relatively weak around its physical value, thus  providing support to the assumption made in Ref.~\cite{Alvarez-Ruso:2013fza} that the SU(2) baryon chiral perturbation theory can be applied to study $n_f=2+1$ lattice QCD simulations as long as the strange quark mass is close to its physical value.
\end{abstract}


\date{\today}

\maketitle

\section{Introduction}

Chiral perturbation theory~(ChPT) provides a model independent framework to explore the nonperturbative regime of strong interactions~\cite{Weinberg:1978kz,Gasser:1983yg,Gasser:1984gg,Gasser:1987rb}.
The formalism and main achievements of ChPT have been reviewed in Refs.~\cite{Leutwyler:1994fi, Bernard:1995dp,Pich:1995bw,Ecker:1994gg,Scherer:2002tk,Bernard:2007zu,Scherer2012b}. As a low-energy effective field theory of quantum chromodynamics (QCD),
it contains a finite number of low energy constants (LECs) up to a certain order, which encode high energy physics integrated out and  can, in principle, only be determined by fitting to experimental data.
The number of unknown LECs becomes large for high order studies, especially in three ($u$, $d$, $s$) flavors, and therefore, practical applications of ChPT are in most cases restricted to low orders.
Fortunately, with the advancement of  numerical algorithms and the continuous increase of computer power, lattice QCD (LQCD) simulations~\cite{Wilson:1974sk} have achieved great success in the study of nonperturbative QCD (see, e.g., Refs.~\cite{Fodor:2012gf,Aoki:2013ldr}) and in addition provided an alternative way to help determine the values of the LECs present in high order chiral Lagrangians.

Recently, several LQCD collaborations have performed fully dynamical simulations with $n_f=2+1$ flavors
for the lowest-lying octet baryon masses~\cite{Durr:2008zz,Aoki:2008sm,WalkerLoud:2008bp, Lin:2008pr,Alexandrou:2009qu,Aoki:2009ix,Bietenholz:2011qq,Beane:2011pc}, which have stimulated many studies of the corresponding chiral extrapolations and the lattice artifacts in ChPT up to next-to-next-to-next-to-leading order (N$^3$LO)~\cite{WalkerLoud:2008bp,Ishikawa:2009vc,Young:2009zb,MartinCamalich:2010fp,Geng:2011wq,Semke:2011ez, Semke:2012gs,Lutz:2012mq,Bruns:2012eh,Ren:2012aj,Ren:2013dzt,Ren:2013wxa,Shanahan:2013cd,Ren:2014vea}.
Because of the large non-vanishing baryon masses in the chiral limit and the resulting power-counting breaking problem~\cite{Gasser:1987rb}, several baryon chiral perturbation theory (BChPT) formulations have been developed, such as heavy baryon~\cite{Jenkins:1990jv}, infrared~\cite{Becher:1999he} and extended-on-mass-shell (EOMS)~\cite{Gegelia:1999gf,Fuchs:2003qc}. Among them, the EOMS approach appears to be phenomenologically successful according to recent studies~\cite{Geng:2008mf, Geng:2010vw, Geng:2010df, Ren:2012aj, Alarcon:2012kn, Chen:2012nx, Blin:2014rpa, Blin:2016itn, Siemens:2016hdi, Yao:2016vbz, Siemens:2016jwj}. Such a success has not been fully understood. In some cases, e.g., for the scalar form factor of the nucleon at $t = 4 m_\pi^2$~\cite{Scherer:2009bt}, it can be attributed to the fact that EOMS is covariant and satisfies analyticity in the loop amplitudes. For other quantities such as the octet baryon masses~\cite{MartinCamalich:2010fp}, the good phenomenological description is somehow unexpected from a power-counting perspective. In Ref.~\cite{Ren:2012aj}, the octet baryon masses have been calculated
up to N$^3$LO in the covariant BChPT with EOMS scheme, and the corresponding $19$ LECs have been determined by a simultaneous fit to the PACS-CS~\cite{Aoki:2008sm}, LHPC~\cite{WalkerLoud:2008bp}, QCDSF-UKQCD~\cite{Bietenholz:2011qq}, HSC~\cite{Lin:2008pr}, and NPLQCD~\cite{Beane:2011pc} lattice data. In order to better constrain the LECs, in Ref.~\cite{Ren:2014vea}, the high statistics lattice data of the PACS-CS~\cite{Aoki:2008sm}, LHPC~\cite{WalkerLoud:2008bp}, QCDSF-UKQCD~\cite{Bietenholz:2011qq} collaborations were reanalyzed, with further constraints provided by the strong isospin breaking effects on the octet baryon masses at the physical point. However, due to the scarcity and limitation of presently available lattice data, it is advisable to exercise caution in using the so-determined LECs to study other related physical quantities.

In Ref.~\cite{Alvarez-Ruso:2013fza},  nucleon masses from the $n_f=2+1$ lattice simulations of BMW~\cite{Durr:2011mp}, PACS-CS~\cite{Aoki:2008sm}, LHPC~\cite{WalkerLoud:2008bp}, HSC~\cite{Lin:2008pr}, NPLQCD~\cite{Beane:2011pc}, MILC~\cite{Aubin:2004wf}, and RBC-UKQCD~\cite{Jung:2012rz} were
analyzed in  SU(2) BChPT with the EOMS scheme as well, with the assumption that the LECs depend only  weakly on the strange quark mass around its physical value. If this assumption holds, because of the relatively faster convergence of SU(2) BChPT in comparison with its SU(3) counterpart, it would in principle provide a more reliable determination of the nucleon mass dependence on the $u/d$ quark masses, and thus the pion-nucleon sigma term via the Feynman-Hellmann theorem.

In the present work, we wish to test the consistency between the SU(3) and SU(2) BChPT descriptions of the nucleon mass by matching the SU(3) BChPT to the SU(2) one.
In particular, we compare certain combinations of the SU(3) LECs with their SU(2) counterparts. This can be
achieved by treating the strange quark mass as a heavy scale compared to the light quark mass and expanding the SU(3) nucleon mass in terms
of  $m_q/m_s$, where $m_q$ is the average $u$ and $d$ quark masses and $m_s$ is the strange quark mass. Since the LECs in Ref.~\cite{Ren:2014vea} and those
in Ref.~\cite{Alvarez-Ruso:2013fza} are determined by fitting to different lattice QCD simulations with varying strategies, the consistency between them will provide a nontrivial check on the validity of the obtained LECs, particularly the SU(3) ones, and on the assumption made in Ref.~\cite{Alvarez-Ruso:2013fza} that the dependence of the SU(2) LECs on the strange quark mass is mild close to the physical point. Furthermore, the relevant pion-nucleon sigma term $\sigma_{\pi N}$ is also evaluated. But one should treat this value with care because none of recent simulations at the physical point Refs.~\cite{Durr:2015dna,Yang:2015uis,Abdel-Rehim:2016won,Bali:2016lvx} were available back when the studies of Refs.~\cite{Ren:2014vea,Alvarez-Ruso:2013fza} were performed.

We note that in Refs.~\cite{Frink:2004ic,Mai:2009ce}, the SU(3) baryon masses and meson-baryon scattering lengths were matched to
their SU(2) counterparts with the aim of constraining the large number of unknown SU(3) LECs with the SU(2) inputs.  In the present case, because of the abundant $n_f=2+1$ LQCD baryon masses, both
the SU(3) and SU(2) LECs have been independently determined in Refs.~\cite{Ren:2014vea,Alvarez-Ruso:2013fza}.  This provides us a unique opportunity to study the flavor dependence of BChPT.~\footnote{In future, the SU(3) BChPT can also be contrasted with the SU(2)  BChPT, e.g., the HB ChPT  of Ref.~\cite{Tiburzi:2008bk},  for hyperon masses once the relevant LECs are fixed in some way.}

This paper is organized as follows. In Sec II, we describe the procedure and strategy used to match the SU(3) nucleon mass to the SU(2) one. Hereby we obtain an effective SU(2) expression for the nucleon mass deduced from the SU(3) one. In Sec. III, we compare the effective SU(2) nucleon mass and pion-nucleon sigma term with the original SU(2) and SU(3) ones, and study the dependence of the SU(2) effective parameters on the strange quark mass. This is followed by a short summary in Sec. IV.

\section{Theoretical Framework}

\begin{figure}[t!]
\centering
\includegraphics[width=13cm]{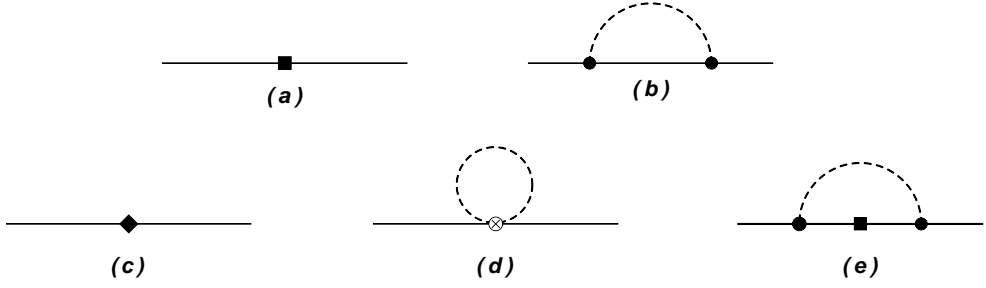}
\caption{Feynman diagrams contributing to the octet-baryon masses up to $\mathcal{O}(p^4)$ in the EOMS-BChPT. The solid lines denote octet-baryons and dashed lines refer to Goldstone bosons.
The black boxes (diamonds) indicate second (fourth) order couplings. The solid dot (circle-cross)
indicates an insertion from the dimension one (two) meson-baryon Lagrangians. Although not explicitly shown, wave-function renormalization is also taken into account and included in $H_{NB}^{(e)}$ of Eq.~(\ref{Eq:SU3}).}
\label{Fig:feydia}
\end{figure}

In this section, we explain in detail how one can match the SU(3) nucleon mass to the SU(2) one by assuming that the strange quark contribution can be integrated out, namely taking $m_q/m_s$ as a small expansion parameter, where $m_q$ is the average $u$ and $d$ quark masses and $m_s$ is the strange quark mass. In the SU(3) EOMS BChPT, the chiral expansion of the nucleon mass up to  $\mathcal{O}(p^4)$ can be written as
\begin{eqnarray}\label{Eq:SU3}
  M_{N}^\mathrm{SU(3)} &=& m_0 + m_N^{(2)} + m_{N}^{(3)} + m_{N}^{(4)} \nonumber\\
  &=& m_0 + \xi_{N\pi}^{(a)}m_\pi^2 + \xi_{NK}^{(a)}m_K^2 + \xi_{N\pi}^{(c)}m_\pi^4 + \xi_{NK}^{(c)}m_K^4 + \xi_{N\pi K}^{(c)}m_\pi^2 m_K^2 \nonumber\\
  && + \frac{1}{(4\pi F_\phi)^2}\sum\limits_{\phi=\pi,~K,~\eta}\left[ \xi_{N\phi}^{(b)} H_N^{(b)} + \xi_{N\phi}^{(d)} H_N^{(d)} +  \sum\limits_{B=N,~\Lambda,~\Sigma}\xi^{(e)}_{NB\phi}  H_{NB}^{(e)}\right],
\end{eqnarray}
where $m_0$ is the baryon mass in the chiral limit while $m_N^{(2)}$, $m_N^{(3)}$, and $m_N^{(4)}$ are the $\mathcal{O}(p^2)$, $\mathcal{O}(p^3)$, and $\mathcal{O}(p^4)$ chiral contributions~\cite{Ren:2012aj}, respectively. The pseudoscalar meson masses are denoted by $m_\phi$ ($\phi=\pi,~K,~\eta$); $F_\phi$ is the pseudoscalar meson decay constant in the chiral limit, which is taken to be $F_\phi=0.0871$ GeV~\cite{Amoros:2001cp}.  Latin characters $a$, $b$, $c$, $d$, $e$ represent the five Feynman diagrams shown in Fig.~\ref{Fig:feydia}.
The $\xi$ coefficients denote combinations of the $19$ LECs ($m_0$, $b_{0,D,F}$, $b_{1,\cdots,8}$, and $d_{1,\cdots,5,7,8}$) appearing in the octet baryon masses up to N$^3$LO. They are given in Tables~1-5 of Ref.~\cite{Ren:2012aj}, where the corresponding loop functions $H$ can also be found. Note that the loop functions $H$ depend on the meson masses (obtained in leading order ChPT), the chiral limit baryon mass $m_0$, and the NLO mass splittings induced by $b_0$, $b_D$, and $b_F$.

It is convenient to isolate the $s\bar s$ contribution to the meson masses by introducing $ m_{s\bar{s}}^2 = 2 B_0 m_s$.
Using the leading order ChPT, the kaon and eta masses can then be expressed as,
\begin{equation}\label{Eq:Ketamass}
  m_K^2 = \frac{1}{2}(m_\pi^2 + m_{s\bar{s}}^2),\quad m_\eta^2 = \frac{1}{3}(m_\pi^2 + 2 m_{s\bar{s}}^2).
\end{equation}
At the physical point, $m_{s\bar{s}}=\sqrt{2m_K^2-m_\pi^2}=683.2$ MeV, where $m_K$ and $m_\pi$ are the isospin averages of the kaon and pion masses.

Now one can approximate the kaon- and eta-loop contributions to the nucleon self-energy ($\Sigma_{K,~\eta}$) by polynomials of the pion mass. Namely, one replaces  $m_{K,~\eta}$ with $m_{\pi,~s\bar{s}}$ and performs a perturbative expansion   in terms of $m_\pi/m_{s\bar{s}}$ up to fourth order,
\begin{equation}\label{Eq:SigmaKetai}
  \Sigma_{K,~\eta}^{(i)} = A_{K,~\eta}^{(i)} + B_{K,~\eta}^{(i)}m_\pi^2 + C_{K,~\eta}^{(i)}m_\pi^4 + \mathcal{O}\left(\frac{m_\pi}{m_{s\bar{s}}}\right)^5,
\end{equation}
where $i$ denotes the different diagrams ($i=a,\cdots,e$); the expansion coefficients ($A_{K,\eta}^{(i)}$, $B_{K,\eta}^{(i)}$, $C_{K,\eta}^{(i)}$) are given in the Appendix. For the pion-cloud contributions of diagram (e), $\Sigma_{\pi}^{(e)}$, because the leading-order correction to the nucleon mass, $m_{N}^{(2)}=-2(2b_0+b_D+b_F)m_\pi^2-2(b_0+b_D-b_F)m_{s\bar{s}}^2$, contains the strange quark contributions,  it should be expanded as well
\begin{equation}\label{Eq:SigmapiE}
  \Sigma_\pi^{(e)} = \frac{3}{64\pi^2 F_\phi^2}(D+F)^2 H_N^{(e)}(m_0, m_\pi, \Delta m_N, \mu) + A_\pi^{(e)} + B_\pi^{(e)} m_\pi^2 + C_\pi^{(e)} m_\pi^4 + \mathcal{O}\left(\frac{m_\pi}{m_{s\bar{s}}}\right)^5,
\end{equation}
where $D$ and $F$ are the axial-vector coupling constants, $\mu$ denotes the renormalization scale, and $\Delta m_N=-2(2b_0+b_D+b_F)m_\pi^2$ is $m_N^{(2)}$ with vanishing strange quark mass. The coefficients, $A_\pi^{(e)}$, $B_\pi^{(e)}$, and $C_\pi^{(e)}$ are given in the Appendix.

Putting all pieces together, we obtain the SU(2) equivalent  nucleon mass,
\begin{eqnarray}\label{Eq:SU3toSU2}
  M_N &=& m_0^\mathrm{eff} - 4c_1^\mathrm{eff} m_\pi^2 + \alpha^\mathrm{eff} m_\pi^4 + \beta^\mathrm{eff} m_\pi^4 \log\frac{\mu^2}{m_\pi^2} \nonumber\\
  && + \frac{1}{(4\pi F_\phi)^2}\frac{3}{2}(D+F)^2
  \left[ H_N^{(b)}(m_0, m_\pi) + \frac{1}{2}H_N^{(e)}(m_0, m_\pi, \Delta m_N, \mu)\right],
\end{eqnarray}
where the tadpole contributions are separated in two terms proportional to $m_\pi^4$ and $m_\pi^4\log(\mu^2/m_\pi^2)$. The corresponding effective parameters, $m_0^\mathrm{eff}$, $c_1^\mathrm{eff}$, $\alpha^\mathrm{eff}$, and $\beta^\mathrm{eff}$ are combinations of the original SU(3) LECs (underlined) and the expansion parameters in Eqs.~\eqref{Eq:SigmaKetai} and ~\eqref{Eq:SigmapiE},
\begin{eqnarray}
  m_0^\mathrm{eff} &=& \underline{m_0} + A_K^{(a)} + A_K^{(b)} + A_\eta^{(b)} + A_K^{(c)} + A_K^{(d)} + A_\eta^{(d)} + A_K^{(e)} + A_\eta^{(e)},  \label{Eq:m0}\\
  c_1^\mathrm{eff} &=& -\frac{1}{4}\left[\underline{\xi_{N\pi}^{(a)}} + B_K^{(a)} + B_K^{(b)} + B_\eta^{(b)} + B_K^{(c)} + B_K^{(d)} + B_\eta^{(d)} + B_{\pi}^{(e)} + B_K^{(e)} + B_\eta^{(e)}\right],\label{Eq:c1}\\
  \alpha^\mathrm{eff} &=& \underline{\xi_{N\pi}^{(c)}} + C_K^{(b)} + C_\eta^{(b)} + C_K^{(c)} + C_\pi^{(d)} +  C_K^{(d)} + C_\eta^{(d)} + C_\pi^{(e)} + C_K^{(e)} + C_\eta^{(e)} ,\label{Eq:alpha}\\
  \beta^\mathrm{eff} &=& D_\pi^{(d)} + D_\pi^{(e)}. \label{Eq:beta}
\end{eqnarray}
These results, when expanded in $1/m_0$, are consistent with those of Refs.~\cite{Frink:2004ic,Mai:2009ce}.

For comparison, the nucleon mass directly obtained in SU(2) BChPT is~\cite{Alvarez-Ruso:2013fza},
\begin{eqnarray}
  M_N^\mathrm{SU(2)} &=& M_0 - 4c_1 m_\pi^2 + \frac{1}{2}\alpha m_\pi^4 \nonumber\\
  && + \frac{1}{(4\pi f_\pi)^2}\frac{3}{8}\left[2(-8c_1+c_2+4c_3)+c_2\right] m_\pi^4 - \frac{1}{(4\pi f_\pi)^2}\frac{3}{4}(8c_1-c_2-4c_3)m_\pi^4\log\frac{\mu^2}{m_\pi^2}\nonumber\\
  && + \frac{1}{(4\pi f_\pi)^2}\frac{3}{2}g_A^2
  \left[ H_N^{(b)}(M_0, m_\pi) + \frac{1}{2}H_N^{(e)}(M_0, m_\pi, (-4c_1 m_\pi^2), \mu)\right],
\end{eqnarray}
where $M_0$ is the nucleon mass in the SU(2) chiral limit with $m_u=m_d=0$ and $m_s$ fixed at its physical value; $c_{1,2,3}$ and $\alpha$ are the unknown LECs. In order to obtain the same form as Eq.~\eqref{Eq:SU3toSU2}, the above equation can be rewritten as
\begin{eqnarray}\label{Eq:SU2}
M_N^\mathrm{SU(2)}&=& M_0 -4c_1 m_\pi^2 + \alpha^\mathrm{SU(2)} m_\pi^4 + \beta^\mathrm{SU(2)} m_\pi^4 \log\frac{\mu^2}{m_\pi^2} \nonumber\\
  && + \frac{1}{(4\pi f_\pi)^2}\frac{3}{2}g_A^2
  \left[ H_N^{(b)}(M_0, m_\pi) + \frac{1}{2}H_N^{(e)}(M_0, m_\pi, (-4c_1 m_\pi^2), \mu)\right],
\end{eqnarray}
with the following two combinations of the LECs,
\begin{eqnarray}
  \alpha^\mathrm{SU(2)} &=& \frac{1}{2}\alpha - \frac{1}{(4\pi f_\pi)^2}\frac{3}{4}\left[(8c_1-c_2-4c_3)-\frac{1}{2}c_2\right],\nonumber\\
  \beta^\mathrm{SU(2)} &=&  - \frac{3}{4(4\pi f_\pi)^2}(8c_1-c_2-4c_3).
\end{eqnarray}

\section{Results and Discussion}

In this section, we evaluate the effective parameters, $m_0^\mathrm{eff}$, $c_1^\mathrm{eff}$, $\alpha^\mathrm{eff}$, and $\beta^\mathrm{eff}$, and compare them with the SU(2) LECs appearing in Eq.~\eqref{Eq:SU2}.
In Ref.~\cite{Ren:2014vea}, the values of the $19$ LECs ($m_0$, $b_{0,D,F}$, $b_{1,\cdots,8}$, and $d_{1,\cdots,5,7,8}$) in the octet baryon masses up to $\mathcal{O}(p^4)$ are determined by fitting the high statistics lattice data of the PACS-CS, LHPC and QCDSF-UKQCD collaborations. In order to better constrain the large number of unknown LECs, the strong isospin breaking effects on the octet baryon masses are also taken into account. As the LQCD data are still limited, it is worthwhile to investigate the consistency of the extracted LECs~\cite{Ren:2014vea}. For this purpose we compare the SU(2) equivalent nucleon mass with the SU(2) one. As a first check, the four combinations of LECs ($m_0^\mathrm{eff}$, $c_1^\mathrm{eff}$, $\alpha^\mathrm{eff}$, $\beta^\mathrm{eff}$) are compared to the SU(2) LECs ($M_0$, $c_1$, $\alpha^\mathrm{SU(2)}$, $\beta^\mathrm{SU(2)}$) of Eq.~\eqref{Eq:SU2}.  In Ref.~\cite{Alvarez-Ruso:2013fza}, these SU(2) LECs have been obtained from the $n_f=2+1$ LQCD data for the nucleon mass, with the strange quark mass close to its physical value. Therefore, they should implicitly incorporate the strange quark contribution that is apparent in Eqs.~(\ref{Eq:m0}-\ref{Eq:beta}).

\begin{table}[ht!]
  \centering
  \caption{Values of the effective parameters after matching the SU(3) nucleon mass to the SU(2) sector [Eq.~\eqref{Eq:SU3toSU2}] and the corresponding LECs of the SU(2) nucleon mass (see Eq.~\eqref{Eq:SU2} and
  Ref.~\cite{Alvarez-Ruso:2013fza}).}
  \label{Tab:effLECs}
\begin{tabular}{l|l}
  \hline\hline
    SU(3)$\rightarrow$SU(2)  & SU(2)  \\
  \hline
  $m_0^\mathrm{eff}=875(10)$ MeV & $M_0=870(3)$ MeV  \\
  $c_1^\mathrm{eff}=-1.07(4)$ GeV$^{-1}$ & $c_1=-1.15(3)$ GeV$^{-1}$ \\
  $\alpha^\mathrm{eff}=4.81(9)$ GeV$^{-3}$ & $\alpha^\mathrm{SU(2)}=6.27(1.98)$ GeV$^{-3}$   \\
  $\beta^\mathrm{eff}=-4.02(20)$ GeV$^{-3}$ & $\beta^\mathrm{SU(2)}=-7.62(93)$ GeV$^{-3}$ \\
  \hline\hline
\end{tabular}
\end{table}
In Table~\ref{Tab:effLECs}, we tabulate the values of the effective parameters appearing in Eq.~\eqref{Eq:SU3toSU2}, with the strange quark mass fixed at its physical value ($m_{s\bar{s}}=683.2$ MeV). For comparison, the corresponding SU(2) LECs, Eq.~\eqref{Eq:SU2} and Ref.~\cite{Alvarez-Ruso:2013fza}, are listed in the second column. We find that $m_0^\mathrm{eff}$ and $c_1^\mathrm{eff}$ agree well with $M_0$ and $c_1$.~\footnote{We note that the value of $c_1$,  and to a less extent, those of $c_2$, $c_3$  (taken from
Ref.~\cite{Alarcon:2011zs}) of Ref.~\cite{Alvarez-Ruso:2013fza}
are consistent with those of Ref.~\cite{Hoferichter:2015hva} within uncertainties.} At $\mathcal{O}(p^4)$, we obtain larger discrepancies: $\alpha^\mathrm{eff}$ is consistent with $\alpha^\mathrm{SU(2)}$ because of the large error bar of the latter; instead, $\beta^\mathrm{eff}$ and $\beta^\mathrm{SU(2)}$ disagree. Although the SU(3) and SU(2) LECs have been obtained with different renormalization scales, $\mu=1$~GeV in Ref.~\cite{Ren:2014vea} and $\mu=M_0$ in Ref.~\cite{Alvarez-Ruso:2013fza}, this only affects the comparison for $\alpha^\mathrm{SU(2)}$, which receives from loop (e) and the $\beta$ term contributions that are small (smaller than the error bar in $\alpha^\mathrm{SU(2)}$ quoted in Table~\ref{Tab:effLECs}). Furthermore, we want to mention that the convergence of the matching looks reasonable where the contributions from $\mathcal{O}(p^2)$, $\mathcal{O}(p^3)$, and $\mathcal{O}(p^4)$ to $c_1$ are $1.10$, $-0.15$, and $0.12$ GeV$^{-1}$, respectively.

\begin{figure}[hb!]
  \centering
  \includegraphics[width=9cm]{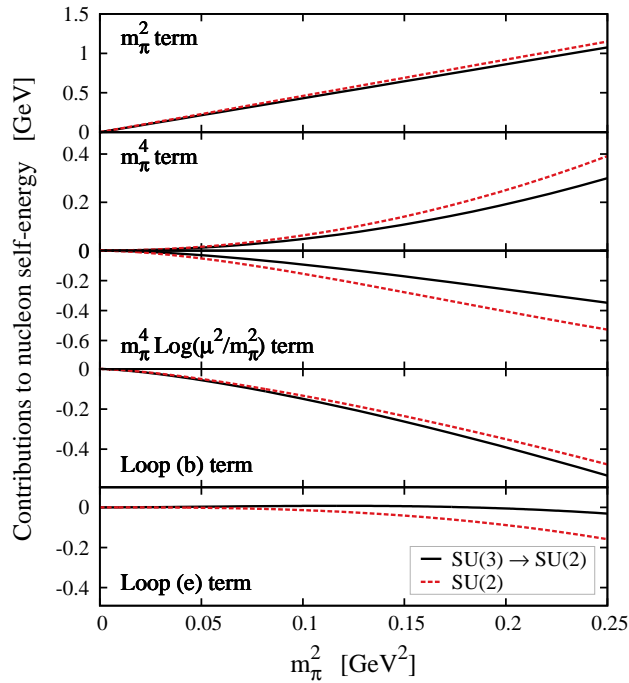}\\
  \caption{Decomposition of the nucleon mass as a function of the pion mass squared (see text for details). The solid lines and the red dashed lines denote the SU(2) equivalent and SU(2) results, respectively.}
  \label{Fig:fiveterms}
\end{figure}

To illustrate the impact of these similarities and differences on the nucleon mass, $m_\pi^2$, $m_\pi^4$ and $m_\pi^4\log(\mu^2/m_\pi^2)$ terms are separately plotted as a function of the leading order $m_\pi^2$  in Fig.~\ref{Fig:fiveterms}. The contributions of the two loop diagrams in Fig.~\ref{Fig:feydia}  are also given. It should be mentioned that the upper limit in the pion mass is set at  $500$ MeV to guarantee a reasonable expansion in powers of $m_\pi/m_{s\bar{s}}$ for $m_{s\bar s}$ close to its physical value.
The agreement is very good for loop (b) and the $m_\pi^2$ term but less so in the rest of terms. This is due to the differences in the central values of $\alpha$ and $\beta$ parameters but also to the above mentioned difference in renormalization scales that reshuffles strength within $\mathcal{O}(p^4)$ terms.

\begin{figure}[t!]
  \centering
  \includegraphics[width=9cm]{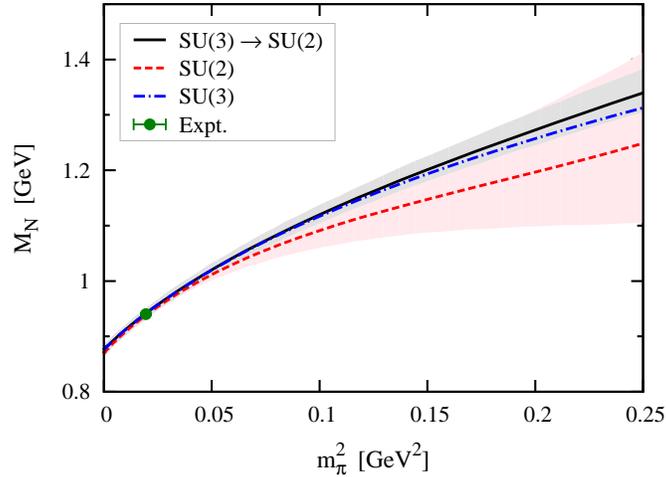}\\
  \caption{Chiral extrapolation of the nucleon mass. The solid line denotes the SU(2) equivalent results, while the red dashed line and the blue dot-dashed line are the SU(2) and SU(3) results, respectively. The green circle denotes the physical point. Error bands for the equivalent SU(2) (narrower) and SU(2) (broader) calculations are also shown.}
  \label{Fig:masschi}
\end{figure}

The pion mass dependence of the nucleon mass for the effective SU(3)~$\rightarrow$~SU(2), SU(2)~\cite{Alvarez-Ruso:2013fza} and SU(3)~\cite{Ren:2014vea} approaches is presented in Fig.~\ref{Fig:masschi}. One can see that the $m_\pi/m_{s\bar{s}}$ expansion truncated at $\mathcal{O}(m_\pi/m_{s\bar{s}})^4$ is a good approximation to the SU(3) case up to rather high $m_\pi^2$. The large error bars in the SU(2) fit make it consistent with both the SU(3) result and the SU(3)~$\rightarrow$~SU(2) projection but there are clear differences in the central values which increase with  $m_\pi^2$. In both Ref.~\cite{Alvarez-Ruso:2013fza} and \cite{Ren:2014vea}, the LQCD pion masses are identified with the next-to-leading order pion masses $M_\pi$ but the way to express $M_N$ in terms of $M_\pi$ is different. In Ref.~\cite{Alvarez-Ruso:2013fza}, higher order terms are neglected by taking $m_N^{(4)}(M_\pi) \approx m_N^{(4)}(m_\pi)$ while in Ref.~\cite{Ren:2014vea} these terms are included by numerically expressing $\mathcal{O}(p^4)$ meson masses in terms of $\mathcal{O}(p^2)$ ones. Although formally equivalent, these two procedures lead to numerically different nucleon masses at high pion masses (for a given set of parameters). However, we have checked that these differences are largely compensated by the different $\mu$ adopted in the two studies: if a given set of LQCD data for the nucleon mass are fitted with Eq.~(\ref{Eq:SU2}) using  $m_N^{(4)}(M_\pi) \approx m_N^{(4)}(m_\pi)$ and $\mu = M_0$ and, on the other hand, applying the numerical inversion of Ref.~\cite{Ren:2014vea} with $\mu = 1$~GeV, the resulting parameters are remarkably close. From this we conclude that the tension between the SU(2) nucleon mass from Ref.~\cite{Alvarez-Ruso:2013fza} and the SU(3) one from Ref.~\cite{Ren:2014vea}, or in the LEC comparison of Table~\ref{Tab:effLECs}, predominantly follows from the use of different data sets, once the SU(3) study incorporates LQCD output for the other octet baryon masses.

\begin{figure}[b!]
  \centering
  \includegraphics[width=9cm]{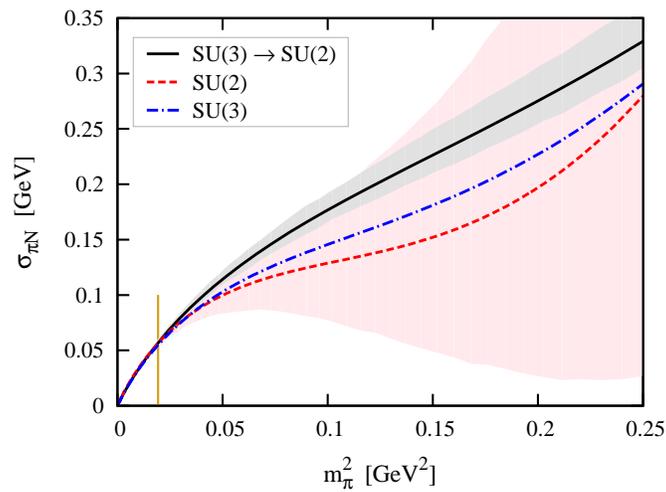}\\
  \caption{Pion mass dependence of the $\sigma_{\pi N}$ term. Line styles are the same as Fig.~\ref{Fig:masschi}, while the vertical orange line denotes the physical pion mass.}
  \label{Fig:sigmampi}
\end{figure}
Nucleon sigma terms play an important role in our understanding of the non-perturbative strong interactions and in searches for beyond standard model physics (see, e.g., Refs.~\cite{Ren:2014vea,Takeda:2010cw,Babich:2010at,Bali:2011ks,Dinter:2012tt,Engelhardt:2012gd,Oksuzian:2012rzb,
Gong:2013vja,Alexandrou:2013nda,Durr:2015dna,Yang:2015uis,Abdel-Rehim:2016won,Bali:2016lvx, Young:2009zb,Oksuzian:2012rzb,Durr:2011mp,Horsley:2011wr,Semke:2012gs,Shanahan:2012wh,Ren:2012aj,
Jung:2012rz,Junnarkar:2013ac} for some recent discussions). One can use the SU(2) equivalent chiral expansion to predict the pion-nucleon sigma term, $\sigma_{\pi N}$, utilizing the Feynman-Hellmann theorem. The obtained result is $\sigma_{\pi N}=57(6)$ MeV at the physical point, which is consistent with the SU(2) and SU(3) values, $\sigma_{\pi N}^\mathrm{SU(2)}=58(3)$ MeV~\cite{Alvarez-Ruso:2013fza} and $\sigma_{\pi N}^\mathrm{SU(3)}=57(2)$ MeV, respectively. \footnote{These
values are consistent with those obtained from the pion-nucleon scattering analysis~\cite{Alarcon:2011zs,Hoferichter:2015dsa,Hoferichter:2016ocj,Yao:2016vbz} but substantially larger than the latest LQCD results~\cite{Durr:2015dna,Yang:2015uis,Abdel-Rehim:2016won,Bali:2016lvx}.
It should be noted that none of the simulations at the physical point were available back when the studies of Refs.~\cite{Ren:2014vea,Alvarez-Ruso:2013fza} were performed. To understand the discrepancy, it is important to take these new results into account. In addition, a careful analysis of the effects of virtual decuplet baryons such as that performed in Ref.~\cite{Ren:2013dzt} might be needed. However, this is
beyond the scope of the present study.} The pion mass dependence of the $\sigma_{\pi N}$ term is shown in Fig.~\ref{Fig:sigmampi}. For higher pion masses, one can see larger differences between the central values than for $M_N$ (Fig.~\ref{Fig:masschi}), particularly between the SU(3) and  SU(3)~$\rightarrow$~SU(2) results, but the consistency is guaranteed by the error bars.

\begin{figure}[t!]
  \centering
  \includegraphics[width=9cm]{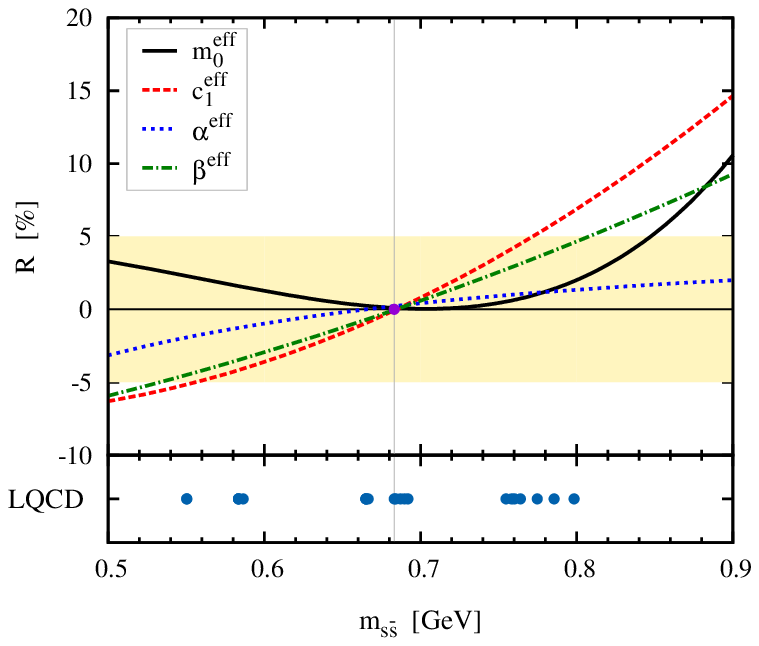}\\
  \caption{Dependence of the relative deviation $R=(X-X^\mathrm{phys.})/X^\mathrm{phys.}$ of the effective parameters on the strange quark mass ($m_{s\bar{s}}=2B_0 m_s$). The black solid line and the red dashed line are the evolutions of the $m_0^\mathrm{eff}$ and $c_1^\mathrm{eff}$, respectively. The blue dotted line and the doted-dashed line represent the results of $\alpha^\mathrm{eff}$ and $\beta^\mathrm{eff}$. The yellow band covers the region of $|R|\leq 5\%$.  The strange quark masses employed in LQCD simulations (BMW, PACS-CS, LHPC, HSC, NPLQCD, MILC, RBC-UKQCD) are presented as blue circles in the lower panel.}
  \label{Fig:strangede}
\end{figure}
\begin{figure}[h!]
  \centering
  \includegraphics[width=9cm]{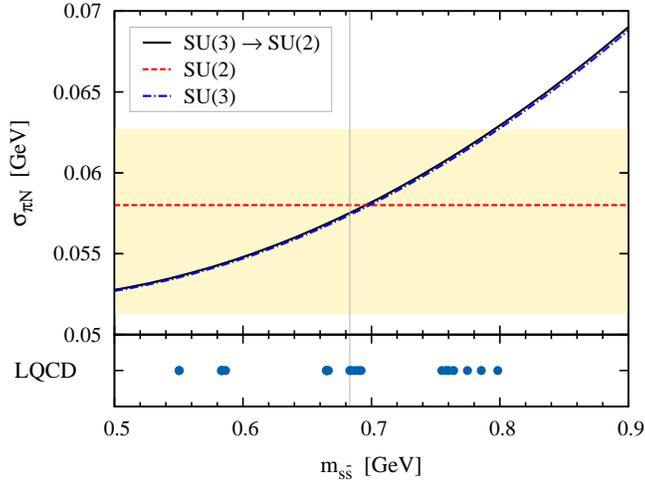}\\
  \caption{Strange quark mass ($m_{s\bar{s}}=2B_0 m_s$) dependence of the $\sigma_{\pi N}$ term. Line styles are the same as in Fig.~\ref{Fig:masschi}. The yellow band indicates a 10\% deviation from the central value of $\sigma_{\pi N}$. As in Fig.~\ref{Fig:strangede}, the lower panel indicates the strange quark masses of different LQCD simulations.}
  \label{Fig:sigmampis}
\end{figure}

As mentioned in the introduction, Ref.~\cite{Alvarez-Ruso:2013fza} reported a global analysis of the $n_f=2+1$ lattice nucleon mass from the BMW~\cite{Durr:2011mp}, PACS-CS~\cite{Aoki:2008sm}, LHPC~\cite{WalkerLoud:2008bp}, HSC~\cite{Lin:2008pr}, NPLQCD~\cite{Beane:2011pc}, MILC~\cite{Aubin:2004wf}, and RBC-UKQCD~\cite{Jung:2012rz} collaborations by using the SU(2) nucleon mass with the assumption that LECs depend weakly on the strange quark mass around its physical value. In Fig.~\ref{Fig:strangede}, the strange quark masses employed in the above LQCD simulations are given in the lower panel. It can be seen that the strange quark mass adopted in the LQCD simulations ($0.55$ GeV $<m_{s\bar{s}}<0.80$ GeV) indeed is close to its physical value , therefore, it is interesting to explore the dependence of the SU(2) equivalent LECs on the strange quark mass. For this, we define the relative deviation $R$ as
\begin{equation}
   R=\frac{X-X^\mathrm{phys.}}{X^\mathrm{phys.}},
\end{equation}
with $X=m_0^\mathrm{eff}, c_1^\mathrm{eff}, \alpha^\mathrm{eff}, \beta^\mathrm{eff}$. In the upper panel of Fig.~\ref{Fig:strangede}, the relative deviation $R$ for the four effective parameters is shown as a function of the strange quark mass. It is observed that the values of the parameters change very little, with $|R|<5\%$, in the range of the strange quark mass employed by LQCD simulations. This study  gives an estimate about the range of the strange quark masses employed in the $n_f=2+1$ LQCD simulations  suitable for an SU(2) BChPT study. It is also interesting to consider the $m_{s\bar{s}}$ dependence of the $\pi N$ sigma term. Figure~\ref{Fig:sigmampis} shows deviations of at most 10\% from the value at the physical point (see the band), in the $m_s$ range of LQCD simulations. Both Figs.~\ref{Fig:strangede}, \ref{Fig:sigmampis} show asymmetries in the slope of some effective LECs and $\sigma_{\pi N}$ above and below the physical $m_{s\bar{s}}$ value. The relatively faster growth of these values could reflect a slower convergence of BChPT for heavier strange quark masses and might introduce biases in SU(2) analyses of $n_f=2+1$  LQCD data.

\section{Conclusion}
We have checked the consistency between the SU(2) and  SU(3) baryon chiral perturbation theory for the nucleon mass.
It is shown that although the number of LECs in the SU(2) and the SU(3) cases is quite different, and
the strategy to fix them using LQCD simulations varies, the so-obtained LECs are largely consistent with each other. In addition,
we have shown that the SU(2) equivalent LECs indeed depend rather weakly on the strange quark mass close to its physical value.
This result further supports the idea that LQCD simulations provide an new alternative way to determine
unknown LECs in baryon chiral perturbation theory, which might be hard to fix otherwise.

With the SU(2) equivalent chiral expansion reported here, we find a $\sigma_{\pi N}=57(6)$ MeV, which is consistent with the results of Refs.~\cite{Ren:2014vea,Alvarez-Ruso:2013fza}. On the other hand, one should take this value with caution, because neither the present study nor Refs.~\cite{Ren:2014vea,Alvarez-Ruso:2013fza} include the latest LQCD simulations at the physical point which appeared after Refs.~\cite{Ren:2014vea,Alvarez-Ruso:2013fza} were published. The current tension between the large sigma term obtained in $\pi$-$N$ scattering analyses and the present study and those of the latest LQCD simulations calls for a new global analysis that includes these physical point lattice data in the fits.
In addition, in Ref.~\cite{Leutwyler:2015jga}, the chiral convergence of $\sigma_0$  was discussed in detail, emphasizing the breakdown of the chiral expansion in case of a large nucleon sigma term. We expect to gain further insight into this issue from a systematic study of all the state of the art LQCD simulations.

\acknowledgements
X.-L.R and L.-S.G. acknowledge the hospitality of Instituto de F\'isica Corpuscular, where this project was initiated. This work is partly supported by the National Natural Science Foundation of China under Grant Nos. 11335002, 11375024, 11522539, 11411130147, and 11621131001, the Fundamental Research Funds for the Central Universities, the China Postdoctoral Science Foundation under Grant No. 2016M600845, the Research Fund for the Doctoral Program of Higher Education under Grant No. 20110001110087. This research has been partially supported by the Spanish Ministerio de Econom\'ia y Competitividad (MINECO) and the European fund for regional development (FEDER) under Contracts FIS2011-28853-C02-01, FIS2011-28853-C02-02, FIS2014-51948-C2-1-P, FIS2014-51948-C2-2-P, SEV-2014-0398 by Generalitat Valenciana under Contract PROMETEOII/2014/0068 and by the European Union HadronPhysics3 project, grant agreement no. 283286.

\section*{Appendix}

In this section, we provide explicitly the expansion coefficients appearing in Eqs.~\eqref{Eq:SigmaKetai} and~\eqref{Eq:SigmapiE}.

\begin{itemize}
\item For diagram (a)
\begin{equation}
  A_{K}^{(a)}= -2(b_0+b_D-b_F)m_{s\bar{s}}^2, \quad B_{K}^{(a)}= -2(b_0+b_D-b_F),\quad   C_{K}^{(a)}= 0.
\end{equation}

\begin{equation}
  A_\eta^{(a)}= B_\eta^{(a)}= C_\eta^{(a)}=0.
\end{equation}

\item For diagram (b)

\begin{eqnarray}
  A_K^{(b)} &=& \frac{m_{s\bar{s}}^3(5D^2-6DF+9F^2)}{384\pi^2 f_\pi^2 m_0} \left[ m_{s\bar{s}}\log\frac{2m_0^2}{m_{s\bar{s}}^2} -2\sqrt{8m_0^2-m_{s\bar{s}}^2}\arccos\frac{m_{s\bar{s}}}{2\sqrt{2}m_0}\right].
\end{eqnarray}

\begin{eqnarray}
  B_K^{(b)} &=& \frac{m_{s\bar{s}}(5D^2-6DF+9F^2)}{192\pi^2f_\pi^2 m_0}  \left[ m_{s\bar{s}} \log\frac{2m_0^2}{m_{s\bar{s}}^2} + \frac{2(m_{s\bar{s}}^2-3m_0^2)}{\sqrt{8m_0^2-m_{s\bar{s}}^2}} \arccos\frac{m_{s\bar{s}}}{2\sqrt{2}m_0}\right].
\end{eqnarray}

\begin{eqnarray}
  C_K^{(b)} &=& \frac{5D^2-6DF+9F^2}{384\pi^2f_\pi^2m_0m_{s\bar{s}}(8m_0^2-m_{s\bar{s}}^2)^{3/2}}
  \left[ -m_{s\bar{s}}\sqrt{8m_0^2-m_{s\bar{s}}^2}\left( (m_{s\bar{s}}^2-8m_0^2)\log\frac{2m_0^2}{m_{s\bar{s}}^2} + 2m_0^2\right) \right.\nonumber\\
  && \left. -2(24m_0^4-12m_0^2m_{s\bar{s}}^2+m_{s\bar{s}}^4) \arccos\frac{m_{s\bar{s}}}{2\sqrt{2}m_0} \right].
\end{eqnarray}

\begin{eqnarray}
  A_\eta^{(b)} &=& \frac{m_{s\bar{s}}^3 (D-3F)^2}{432\pi^2f_\pi^2 m_0} \left[ m_{s\bar{s}}\log\frac{3m_0^2}{2m_{s\bar{s}}^2} - 2\sqrt{6m_0^2-m_{s\bar{s}}^2}\arccos\frac{m_{s\bar{s}}}{\sqrt{6}m_0}\right].
\end{eqnarray}

\begin{eqnarray}
  B_\eta^{(b)} &=& \frac{m_{s\bar{s}}(D-3F)^2}{432\pi^2f_\pi^2m_0}  \left[ m_{s\bar{s}} \log\frac{3m_0^2}{2m_{s\bar{s}}^2} + \frac{2m_{s\bar{s}}^2-9m_0^2}{\sqrt{6m_0^2-m_{s\bar{s}}^2}} \arccos\frac{m_{s\bar{s}}}{\sqrt{6}m_0}\right].
\end{eqnarray}

\begin{eqnarray}
  C_\eta^{(b)} &= &  -\frac{(D-3F)^2}{3456\pi^2f_\pi^2 m_0 m_{s\bar{s}} (6m_0^2-m_{s\bar{s}}^2)^{3/2}} \left[ m_{s\bar{s}}\sqrt{6m_0^2-m_{s\bar{s}}^2} \left( 2 (m_{s\bar{s}}^2-6m_0^2)  \log\frac{3m_0^2}{2m_{s\bar{s}}^2} + 3m_0^2\right) \right.\nonumber\\
  && \left. + (54m_0^4-36m_0^2m_{s\bar{s}}^2 +4m_{s\bar{s}}^4) \arccos\frac{m_{s\bar{s}}}{\sqrt{6}m_0} \right].
\end{eqnarray}

\item For diagram (c)
\begin{eqnarray}
  A_{K}^{(c)} &=& -4(d_1-d_2+d_3-d_5+d_7+d_8)m_{s\bar{s}}^4, \quad B_K^{(c)}=4(2d_1-2d_3+d_5-4d_7)m_{s\bar{s}}^2,\nonumber\\
  C_K^{(c)} &=& 4(3 d_1-d_2-d_3-3 d_7+d_8).
\end{eqnarray}

\begin{equation}
  A_{\eta}^{(c)}= B_\eta^{(c)}=C_\eta^{(c)}=0.
\end{equation}

\item For diagram (d)

\begin{equation}
  C_\pi^{(d)} = -\frac{3}{4(4\pi F_\phi)^2} \left[4(2b_0+b_D+b_F) - 4(b_1+b_2+b_3+2b_4) - 3m_0(b_5+b_6+b_7+2b_8)\right].
\end{equation}

\begin{equation}
  D_\pi^{(d)} = -\frac{3}{2(4\pi F_\phi)^2} \left[2(2b_0+b_D+b_F) - 2(b_1+b_2+b_3+2b_4) - m_0(b_5+b_6+b_7+2b_8)\right].
\end{equation}

\begin{eqnarray}
  A_{K}^{(d)} &=& \frac{1}{128\pi^2f_\pi^2} m_{s\bar{s}}^4 \left[4(-4b_0-3b_D+b_F +3b_1+3b_2-b_3+4b_4)\left(1+\log\frac{2\mu^2}{m_{s\bar{s}}^2}\right)\right.\nonumber\\
  && \left. +m_0(3b_5-b_6+3b_7+4b_8)\left(3+2\log\frac{2\mu^2}{m_{s\bar{s}}^2}\right)\right].
\end{eqnarray}

\begin{eqnarray}
  B_{K}^{(d)} &=& \frac{1}{32\pi^2f_\pi^2}m_{s\bar{s}}^2 \left[(-4b_0-3b_D+b_F+3b_1+3b_2-b_3+4b_4) \left(1+2\log\frac{2\mu^2}{m_{s\bar{s}}^2}\right)\right.\nonumber\\
  && \left.+m_0(3b_5-b_6+3b_7+4b_8)\left(1+\log\frac{2\mu^2}{m_{s\bar{s}}^2}\right)\right].
\end{eqnarray}

\begin{eqnarray}
  C_{K}^{(d)} &=& \frac{1}{64\pi^2f_\pi^2}\left[ (4b_0+3b_D-b_F- 3b_1-3b_2+b_3-4b_4)\left(1-2\log\frac{2\mu^2}{m_{s\bar{s}}^2}\right)\right.\nonumber\\
  && \left. + m_0(3b_5-b_6+3b_7+4b_8)\log\frac{2\mu^2}{m_{s\bar{s}}^2}\right].
\end{eqnarray}

\begin{eqnarray}
  A_\eta^{(d)} &=& \frac{1}{432\pi^2f_\pi^2}m_{s\bar{s}}^4 \left[ (-24(b_0+b_D-b_F) + 4(9b_1+b_2-3b_3+6b_4)) \left(1+\log\frac{3\mu^2}{2m_{s\bar{s}}^2}\right)\right. \nonumber\\
  && \left. + m_0(9b_5-3b_6+b_7+6b_8)\left(3+2\log\frac{3\mu^2}{2m_{s\bar{s}}^2}\right)\right].
\end{eqnarray}

\begin{eqnarray}
  B_\eta^{(d)} &=& \frac{1}{216\pi^2f_\pi^2}m_{s\bar{s}}^2 \left[ (-3(2b_0+b_D+b_F) + 9b_1+b_2-3b_3+6b_4) \left(1+2\log\frac{3\mu^2}{2m_{s\bar{s}}^2}\right)\right.\nonumber\\
  && \left.  + m_0(9b_5-3b_6+b_7+6b_8)\left(1+\log\frac{3\mu^2}{2m_{s\bar{s}}^2}\right) +3(-b_D+3b_F)\log\frac{3\mu^2}{2m_{s\bar{s}}^2}\right].
\end{eqnarray}

\begin{eqnarray}
  C_\eta^{(d)} &=& \frac{1}{864\pi^2f_\pi^2} \left[ (6(b_0+b_D-b_F) -9b_1-b_2+3b_3-6b_4) \left(1-2\log\frac{3\mu^2}{2m_{s\bar{s}}^2}\right)\right.\nonumber\\
  && \left. + (6(b_D-3b_F) + m_0(9b_5-3b_6+b_7+6b_8)) \log\frac{3\mu^2}{2m_{s\bar{s}}^2}\right].
\end{eqnarray}

\item For diagram (e)

\begin{equation}
  A_\pi^{(e)} = 0.
\end{equation}

\begin{equation}
  B_\pi^{(e)} = \frac{3}{16\pi^2f_\pi^2} m_{s\bar{s}}^2(D+F)^2 (b_0+b_D-b_F) \log\frac{m_0^2}{\mu^2}.
\end{equation}

\begin{equation}
  C_\pi^{(e)} = \frac{3}{32\pi^2f_\pi^2 m_0^2} m_{s\bar{s}}^2(D+F)^2 (b_0+b_D-b_F) \log\frac{m_0^2}{\mu^2}.
\end{equation}

\begin{equation}
  D_\pi^{(e)} = \frac{3m_{s\bar{s}}^2}{2(4\pi f_\pi)^2 m_0^2}(D+F)^2 (b_0+b_D-b_F).
\end{equation}

\begin{eqnarray}
A_K^{(e)}&=&
\frac{m_{s\bar{s}}^4(D+3F)^2}{576\pi^2f_\pi^2 m_0^2}
\left[ m_0^2\left(-3(b_D+b_F) + 2(6b_0+7b_D-3b_F)\log\frac{m_0^2}{\mu^2}\right)\right.\nonumber\\
&&\quad \left.+ \frac{1}{2}\left(m_{s\bar{s}}^2(3b_0+2b_D-6b_F)+6m_0^2(b_D+3b_F)\right)\log\frac{2m_0^2}{m_{s\bar{s}}^2}\right.\nonumber\\
&&\left. + \frac{m_{s\bar{s}}}{\sqrt{8m_0^2-m_{s\bar{s}}^2}}\left(m_{s\bar{s}}^2(3b_0+2b_D-6b_F)+12m_0^2(b_D+3b_F)\right)\arccos\frac{m_{s\bar{s}}}{2\sqrt{2}m_0}
\right]\nonumber\\
&+&
\frac{3m_{s\bar{s}}^4(D-F)^2}{64\pi^2f_\pi^2 m_0^2}
\left[ m_0^2\left(3(b_D-b_F) + (2b_0+b_D-b_F)\log\frac{m_0^2}{\mu^2}\right)\right.\nonumber\\
&&\quad\left.+ \frac{1}{2}\left(m_{s\bar{s}}^2(b_0+2b_D-2b_F)-6m_0^2(b_D-b_F)\right)\log\frac{2m_0^2}{m_{s\bar{s}}^2}\right.\nonumber\\
&&\left. + \frac{m_{s\bar{s}}}{\sqrt{8m_0^2-m_{s\bar{s}}^2}}
\left(m_{s\bar{s}}^2(b_0+2b_D-2b_F)-12m_0^2(b_D-b_F)\right)\arccos\frac{m_{s\bar{s}}}{2\sqrt{2}m_0}\right].
\end{eqnarray}

\begin{eqnarray}
B_K^{(e)}&=& \frac{m_{s\bar{s}}^2(D+3F)^2}{576\pi^2f_\pi^2m_0^2}\left[
\frac{m_0^2}{8m_0^2-m_{s\bar{s}}^2}\left(m_{s\bar{s}}^2(-12b_0-11b_D+15b_F) - 24 m_0^2(b_D+3b_F)\right)\right]\nonumber\\
&& + 6m_0^2(3b_0+2b_D)\log\frac{m_0^2}{\mu^2}
+ m_{s\bar{s}}^2(6b_0+4b_D-3b_F)\log\frac{2m_0^2}{m_{s\bar{s}}^2}\nonumber\\
&& + \frac{2m_{s\bar{s}}}{(8m_0^2-m_{s\bar{s}}^2)^{3/2}}
\left(18m_0^2m_{s\bar{s}}^2(3b_0+2b_D-2b_F)-m_{s\bar{s}}^4(6b_0+4b_D-3b_F)\right.\nonumber\\
&&\quad \left.\left.+24m_0^4(b_D+3b_F)\right)
\arccos\frac{m_{s\bar{s}}}{2\sqrt{2}m_0}\right]\nonumber\\
&+& \frac{3m_{s\bar{s}}^2(D-F)^2}{64\pi^2f_\pi^2m_0^2}
\left[
\frac{m_0^2}{8m_0^2-m_{s\bar{s}}^2}\left(m_{s\bar{s}}^2(-4b_0-5b_D+5b_F) + 24 m_0^2(b_D-b_F)\right)\right.\nonumber\\
&&+ 2m_0^2(3b_0+2b_D)\log\frac{m_0^2}{\mu^2}
+ m_{s\bar{s}}^2(2b_0+2b_D-b_F)\log\frac{2m_0^2}{m_{s\bar{s}}^2}\nonumber\\
&&+ \frac{2m_{s\bar{s}}}{(8m_0^2-m_{s\bar{s}}^2)^{3/2}}
\left(2m_0^2m_{s\bar{s}}^2(9b_0+10b_D-6b_F)-m_{s\bar{s}}^4(2b_0+2b_D-b_F)\right.\nonumber\\
&&\quad \left.\left.-24m_0^4(b_D-b_F)\right)
\arccos\frac{m_{s\bar{s}}}{2\sqrt{2}m_0}\right].
\end{eqnarray}

\begin{eqnarray}
C_K^{(e)}&=&
\frac{(D+3F)^2}{1152\pi^2f_\pi^2m_0^2}\left[
\frac{m_0^2}{(8m_0^2-m_{s\bar{s}}^2)^2}\left(m_{s\bar{s}}^4(66b_0+47b_D+21b_F) \right.\right.\nonumber\\
&&\quad \left.\left.- 8 m_{s\bar{s}}^2m_0^2(84b_0+77b_D+39b_F)+576m_0^4(b_D+3b_F)\right)\right.\nonumber\\
&&+ 2m_0^2(12b_0+5b_D+3b_F)\log\frac{m_0^2}{\mu^2}
+ \left(m_{s\bar{s}}^2(15b_0+10b_D+6b_F)-6m_0^2(b_D+3b_F)\right)\log\frac{2m_0^2}{m_{s\bar{s}}^2}\nonumber\\
&&\left.+ \frac{2m_{s\bar{s}}}{(8m_0^2-m_{s\bar{s}}^2)^{5/2}}
\left(m_{s\bar{s}}^6(15b_0+10b_D+6b_F)-4m_{s\bar{s}}^4m_0^2(69b_0+49b_D+33b_F)\right.\right.\nonumber\\
&&\quad \left.\left.+40m_{s\bar{s}}^2m_0^4(33b_0+28b_D+24b_F)
-864m_0^6(b_D+3b_F)\right)\arccos\frac{m_{s\bar{s}}}{2\sqrt{2}m_0}\right]\nonumber\\
&+&
\frac{3(D-F)^2}{128\pi^2f_\pi^2m_0^2}\left[
\frac{m_0^2}{(8m_0^2-m_{s\bar{s}}^2)^2}\left(m_{s\bar{s}}^4(22b_0+9b_D+7b_F) \right.\right.\nonumber\\
&&\quad \left.\left.- 8 m_{s\bar{s}}^2m_0^2(28b_0+3b_D+13b_F)-576m_0^4(b_D-b_F)\right)\right.\nonumber\\
&&+ 2m_0^2(4b_0+3b_D+b_F)\log\frac{m_0^2}{\mu^2}
+ \left(m_{s\bar{s}}^2(5b_0+2b_D+2b_F)+6m_0^2(b_D-b_F)\right)\log\frac{2m_0^2}{m_{s\bar{s}}^2}\nonumber\\
&&\left.+ \frac{2m_{s\bar{s}}}{(8m_0^2-m_{s\bar{s}}^2)^{5/2}}
\left(m_{s\bar{s}}^6(5b_0+2b_D+2b_F)-4m_{s\bar{s}}^4m_0^2(23b_0+7b_D+11b_F)\right.\right.\nonumber\\
&&\quad \left.\left.
+40m_{s\bar{s}}^2m_0^4(11b_0+8b_F)
+864m_0^6(b_D-b_F)\right)\arccos\frac{m_{s\bar{s}}}{2\sqrt{2}m_0}\right].
\end{eqnarray}

\begin{equation}
A_\eta^{(e)}=\frac{m_{s\bar{s}}^4(D-3F)^2(b_0+b_D-bF)}{216\pi^2f_\pi^2 m_0^2}
\left[3m_0^2\log\frac{m_0^2}{\mu^2}+m_{s\bar{s}}^2\log\frac{3m_0^2}{2m_{s\bar{s}}^2} + \frac{2m_{s\bar{s}}^3}{\sqrt{6m_0^2-m_{s\bar{s}}^2}}\arccos\frac{m_{s\bar{s}}}{\sqrt{6}m_0}\right].
\end{equation}

\begin{eqnarray}
B_\eta^{(e)}&=&\frac{m_{s\bar{s}}^2(D-3F)^2}{216\pi^2f_\pi^2 m_0^2}
\left[
-\frac{3m_0^2m_{s\bar{s}}^2(b_0+b_D-b_F)}{6m_0^2-m_{s\bar{s}}^2} +
\frac{3}{2}m_0^2(5b_0+3b_D+b_F)\log\frac{m_0^2}{\mu^2} \right.\nonumber\\
&&+m_{s\bar{s}}^2(3b_0+2b_D)\log\frac{3m_0^2}{2m_{s\bar{s}}^2} \nonumber\\
&&+\left.
\frac{m_{s\bar{s}}^3(3m_0^2(13b_0+9b_D-b_F)-2m_{s\bar{s}}^2(3b_0+2b_D))}
{(6m_0^2-m_{s\bar{s}}^2)^{3/2}}\arccos\frac{m_{s\bar{s}}}{\sqrt{6}m_0}\right].
\end{eqnarray}

\begin{eqnarray}
C_\eta^{(e)}&=&\frac{(D-3F)^2}{1728\pi^2f_\pi^2 m_0^2}
\left[
\frac{m_{s\bar{s}}^2}{(6m_0^2-m_{s\bar{s}}^2)^2}(-36m_0^4(11b_0+7b_D+b_F) + 3m_0^2m_{s\bar{s}}^2(19b_0+11b_D+5b_F)) \right.\nonumber\\
&&+
12m_0^2(2b_0+b_D+b_F)\log\frac{m_0^2}{\mu^2} +
2m_{s\bar{s}}^2(9b_0+5b_D+3b_F)\log\frac{3m_0^2}{2m_{s\bar{s}}^2} \nonumber\\
&&+
\frac{2m_{s\bar{s}}^3}{(6m_0^2-m_{s\bar{s}}^2)^{5/2}}\left(45m_0^4(19b_0+11b_D+5b_F)-6m_0^2m_{s\bar{s}}^2(41b_0+23b_D+13b_F)\right)\nonumber\\
&&\left.\quad\left.+2m_{s\bar{s}}^4(9b_0+5b_D+3b_F)\right)\arccos\frac{m_{s\bar{s}}}{\sqrt{6}m_0}\right].
\end{eqnarray}

\end{itemize}

\bibliographystyle{apsrev4-1}
\bibliography{SU3to2}

\end{document}